\documentclass[smallextended]{svjour3}       % onecolumn (second format)
\smartqed  % flush right qed marks, e.g. at end of proof

\usepackage{lineno}

\usepackage[utf8]{inputenc} %unicode support
\usepackage{subfig}
\usepackage{graphicx,url}
\usepackage{booktabs}
\usepackage{epstopdf}
\usepackage{epsfig}
\usepackage{multicol}
\usepackage{xspace}
\usepackage{amsmath}

\newcommand{\figwidthW}{3.5in}

\newcommand{\grockit}{grockit.com}
\usepackage{color}
\newcommand{\remove}[1]{}
\usepackage[normalem]{ulem} % for \sout
\newcommand{\deleted}[1]{}
\newcommand{\added}[1]{#1}

\begin{document}

\title{Model of Cognitive Dynamics Predicts Performance on Standardized Tests}
\author{Nathan O. Hodas         \and
        Jacob Hunter \and Stephen J. Young
        \and Kristina Lerman\footnotemark[1]
}

%Shaded regions shows prediction errors of model fits
\institute{
            N O Hodas \at
            Pacific Northwest National Lab, %Richland, WA 99354, USA,
             \email{nathan.hodas@pnnl.gov}
            \and
           J Hunter \at
              Pacific Northwest National Lab, %Richland, WA 99354, USA,
              \email{jacob.hunter@pnnl.gov} %\\
%             \emph{Present address:} Carnegie Mellon University
        \and
        S J Young \at
              Pacific Northwest National Lab, %Richland, WA 99354, USA,
              \email{stephen.young@pnnl.gov} %\\
 %            \emph{Present address:} Carnegie Mellon University
            \and
           K. Lerman \at
              USC Information Sciences Institute,
              \email{lerman@isi.edu}           %  \\
}

% Use the asterisk to denote corresponding authorship and provide email address in note below.
%* nhodas@pnnl.gov

\maketitle

\footnotetext[1]{This is a pre-print of an article published in \emph{Journal of Computational Social Science}. The final authenticated version is available online at: https://doi.org/10.1007/s42001-018-0025-x}

% Please keep the abstract below 200 words
\begin{abstract}
In the modern knowledge economy, success demands sustained focus and high cognitive performance. Research suggests that human cognition is linked to a finite resource, and upon its depletion, cognitive functions such as self-control and decision-making may decline. While fatigue, among other factors, affects human activity, how cognitive performance evolves during extended periods of focus remains poorly understood. By analyzing performance of a large cohort answering practice standardized test questions online, we show that accuracy and learning decline as the test session progresses and recover following prolonged breaks. To explain these findings, we hypothesize that answering questions consumes some finite cognitive resources on which performance depends, but these resources recover during breaks between test questions. We propose a dynamic mechanism of the consumption and recovery of these resources and show that it  explains empirical findings and predicts performance better than alternative hypotheses. While further controlled experiments are needed to identify the physiological origin of these phenomena, our work highlights the potential of empirical analysis of large-scale human behavior data to explore cognitive behavior.

\end{abstract}

\section{Introduction}
A growing body of evidence suggests that human cognitive performance in decision making~\cite{danziger2011extraneous,Shah:2012de}, visual attention tasks~\cite{Boksem05,Killeen:2013iw,Healy04}, self-control~\cite{Muraven:1998ta} and even morality~\cite{Morning-morality-effect} declines on relatively short time scales of hours and even minutes. %may be limited by finite resources available to the brain.
Recent studies of online activity demonstrated similar deterioration in performance. For  example, among the comments posted on a social media site over the course of a single session, % characterized as a period of activity without an extended break,
or answers to questions posted on a question--answering forum, those written later in a session have lower quality: they are shorter, less complex, and receive less feedback from others~\cite{Singer2016plosone,Ferrara2017dynamics}. People also prefer to engage in easier tasks later in a session, e.g., skimming posts, rather than reading them~\cite{KootiA2017www} or retweeting the messages of others, rather than composing original messages~\cite{Kooti2016socinfo}.
One hypothesis advanced to explain these findings is that cognitive performance is limited by finite resources available to the brain. According to this view, the brain uses some energetic resources for mental work, and upon depletion of these resources, performance declines.
Laboratory studies have linked impulsive behavior, the loss of willpower and executive function (`ego depletion'), with the consumption of a finite resource~\cite{Muraven:1998ta,Baumeister:2008ge}, believed to be glucose~\cite{Gailliot:2007dr,Gailliot:2007fv}. However, these works have been controversial, and alternative hypotheses exist~\cite{Boksem08,Beedie:2012di,Kurzban:2013cv,Inzlicht:2014kf}, including boredom and brain's strategic choice to limit effort.
The controversy stems in part from the difficulties of measuring cognitive depletion and replicating its effects in laboratory experiments~\cite{Lange:2014iv} and in part from the lack of clear mechanisms to characterize the depletion process.

In the current paper we %clarify
explore the link between cognitive resources and performance
%explore the cognitive depletion hypothesis
through mathematical analysis of large-scale behavioral data.
Specifically, we study performance on online practice standardized tests (SAT, ACT, and GED) collected from a large cohort under natural conditions.
%We demonstrate %deterioration in performance
%that 1) performance deteriorates over the course of a study session, and that 2) this trend is best explained by the depletion of finite cognitive resources.
The data we study, obtained from \grockit, contain records of 2.8 million attempts by 180 thousand users to answer six thousand questions. The data include the time a user started and stopped answering each question, the question's subject matter and outcome, i.e., whether the answer was correct, incorrect, or skipped. Because of the importance of standardized testing in determining student's educational opportunities, understanding how to characterize and maximize performance is an interesting question on its own right. In this paper, however, we focus on the dynamics of test performance. Whereas a previous study showed that time-of-day affects test performance~\cite{sievertsen2016cognitive}, here we demonstrate that the mere act of answering test questions impairs performance and the ability to learn correct answers.
In addition, we show that while performance declines over the course of a test-taking session,
%and there is a time-dependent performance gain after a break between answering questions.
it recovers following prolonged breaks between sessions.

We argue that these empirical observations are best understood if answering questions consumed some finite cognitive resources on which performance depends.
%We developed a kinetic model of cognitive resources, which associates diminished performance with the depletion of cognitive resources.
%%We demonstrate, however, that the model in which the primary driver of cognitive performance is drawn from a finite secondary resource better explains a variety of performance metrics than the model with a single cognitive resource.
We propose a dynamic model of resource depletion
that is inspired by mechanisms for energy metabolism in the brain,
and show that it better explains performance than alternate models. It also provides the foundation for estimating resource levels and predicting performance, and eventually even developing strategies that optimize performance.

Modeling behavioral data presents many challenges, including size, noise and individual variability. We partly account for this variability by rescaling individual parameters
and estimating them using a model fitting procedure. Model fitting maximizes the explanatory power of the model, which is measured non-parameterically by mutual information between model variables and  behavioral outcomes.

The growing availability of behavioral data has opened the psychological and cognitive sciences to new lines of inquiry from the computational perspective. The current paper's contribution lies in illustrating some of the challenges posed by  working with big behavioral data, as well as the opportunities the data offer for the new field of computational cognitive science. %studying dynamics of cognitive performance.

\section{Results}
\begin{figure}[htb] %  figure placement: here, top, bottom, or page
\centering
\includegraphics[width=6in]{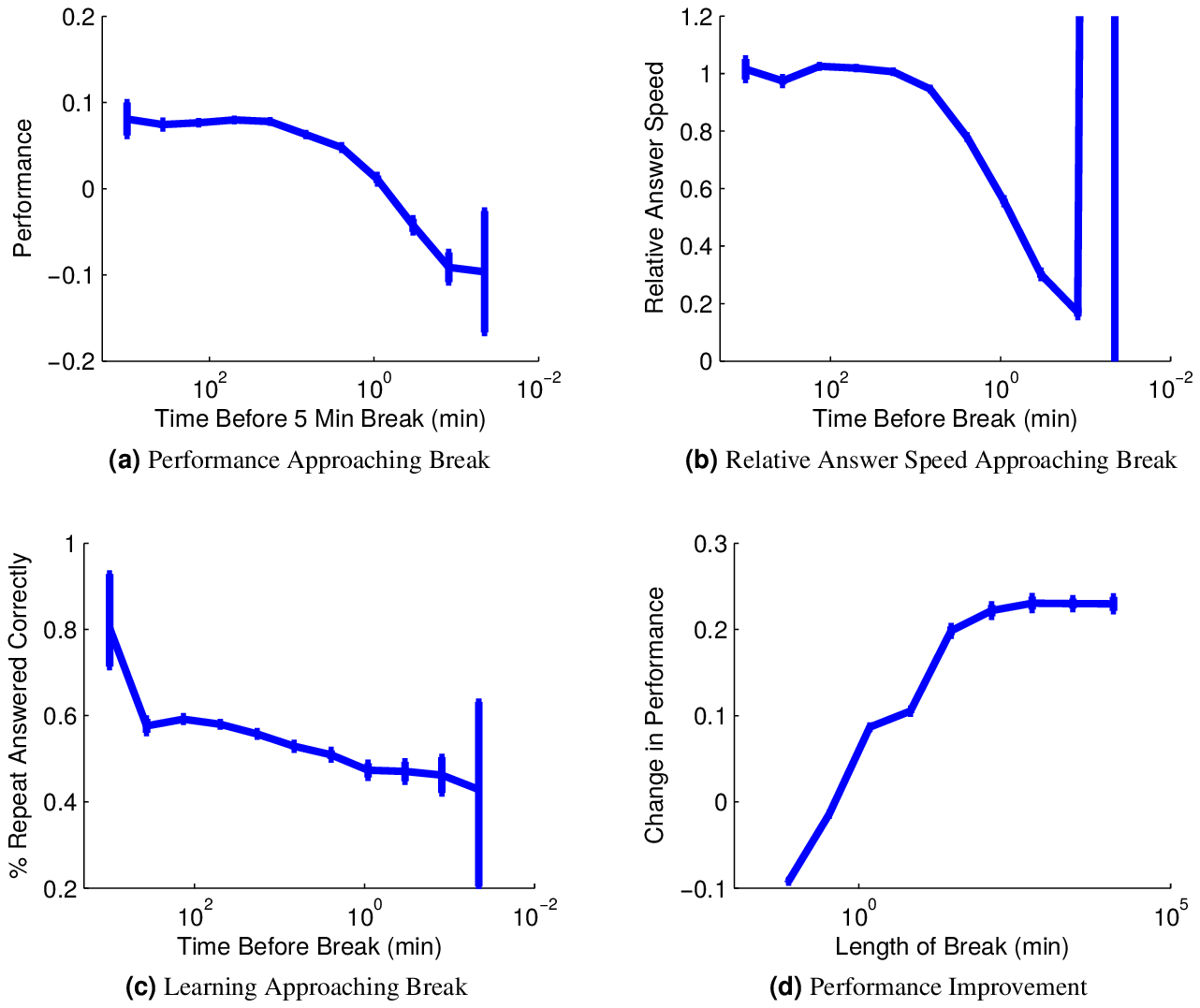}
\subfloat{\label{fig:perfrelbreak}}
\subfloat{\label{fig:ansrelbreak}}
\subfloat{\label{fig:learningbeforebreak}}
\subfloat{\label{fig:perfafterbreak}}
 \caption{\textbf{Observed relationships between performance and time-to-break and length-of-break.}
 Individual plots show that (a) relative accuracy of answers, (b) relative answer speed, and (c) learning all decline as users approach the time they decide to take a break.  On the other hand, (d) performance improves more following longer breaks.
 Note that (b) shows only part of the range on the $y$-axis. Error bars represent standard errors.
 }
\end{figure}

One of the main challenges in analyzing behavioral data is extreme individual heterogeneity. It is necessary to account for this heterogeneity to remove excessive bias and variance from analysis~\cite{Vaupel85heterogeneity,Lloyd-Smith05,Hodas14srep}. Properly characterizing variation between people requires sample sizes far larger than those typically available in a laboratory setting, making real-world online data  an exciting and valuable tool for studying cognitive performance. Without such data, important trends are often obscured. For example, to test for performance deterioration, we might simply track user performance on the practice test over time, but this reveals no trend.% (Figure~\ref{fig:perfovertime}).
Although we expect all users to begin working on practice tests under different initial conditions, for example, whether well-rested or having eaten, they may stop working for similar reasons, such as fatigue, boredom, etc. Instead of aligning the time series of users based on the time they started working, we align them based on the time they decided to stop working and take a break, where we define a break as a period of at least 5 minutes without answering questions (we find no notable variation in effect size for breaks longer than 5 minutes).
Using this alignment, we see a systematic decrease in performance approaching a break (Figure~\ref{fig:perfrelbreak}), even after controlling for user ability and question difficulty as described below.
Note that this occurs even if the users are not aware when they will take a break,
as the last question before the break is often never completed.

To partially account for individual heterogeneity, we first calculate the probability $P(u,q)$ that a user with net accuracy $u$ correctly answers a question with net difficulty $q$ (\% correct answers). Notice that when a question is difficult, $q$ has a low value, meaning that few users have answered it correctly. User's \emph{performance} on an attempt $a$ is $\mathcal{P} = \delta_a - P(u,q)$, where $\delta_a = 1$ if the user answered the question correctly, and 0 otherwise.  In this way, we can compare the outcomes of different users answering different questions. An average performance of $\langle\mathcal{P}\rangle=0$ indicates a user answers questions correctly at the same rate as we expect for those user/question combinations.  Negative performance ($\mathcal{P} < 0$) signifies a user is under-performing, and positive performance ($\mathcal{P} > 0$) signifies a user answering questions correctly more frequently than expected. As users approach a break, their performance decreases (Figure~\ref{fig:perfrelbreak}) and the speed at which they answer questions also decreases until very near break time (Figure~\ref{fig:ansrelbreak}).

After answering a question on {\grockit}, the user is presented with the correct answer, regardless of the outcome, and thus he or she has the opportunity to learn both the solution method and the precise answer.  If the user is presented  the same question again, it will have the same answer, since each question/answer pair has a unique id, so questions with the same id will have the same answer.   To estimate learning, we search for the next question with the same question-id attempted by the user. We limited the calculation only to questions the user actually provided a response, so they had the opportunity to be exposed to the correct answer. If users learned (or at least remembered) the questions from previous attempts, they should be able to answer them correctly upon repeat exposure.  However, we find a systematic decline  in users' %ability to recall the correct answer
accuracy as a function of time before a break they were exposed to the correct answer (Figure~\ref{fig:learningbeforebreak}).

Conversely, the length of the break is highly correlated with performance.  As the time from the end of one question to the start of the next question increases, the user's relative performance  increases (Figure~\ref{fig:perfafterbreak}).  That is, while the absolute performance does not vary significantly as a function of time between questions (or may even decrease due to ``warm-up'' requirements described below), the relative change in performance from the previous question to the next   one  increases with the between-question time interval.
This suggests that longer breaks between questions are associated with under-performance, and the worse the under-performance, the longer the break the user takes, recovering the performance after the break.
The same trend holds for learning: users appear to take longer breaks after their ability to learn the answer has decreased (not shown).

To understand these observations and predict performance, we developed two kinetic models of cognitive resource depletion that are inspired by recent neuroenergetics models~\cite{Killeen:2013iw}.
The first model, the ``one-resource" model, was motivated by work exploring the link between glucose and finite willpower~\cite{Muraven:1998ta,Gailliot:2007fv,Gailliot:2007dr}. The model %, specified by Eq.~\ref{eq:1resource},
considers a single resource $A$,  that decreases while the user is working (i.e., attempting to answer a question) and recovers during periods of rest (i.e., time interval between question-answer attempts). By implication, $A$ is the primary driver of performance~\cite{Killeen:2013iw}.
The general form for the one-resource model is
\begin{equation}\label{eq:A}
\dot{A}(t) = -w_1(A,t)\delta(t) + r_1(A,t) (1-\delta(t)),
\end{equation}
where the functions $w_1$ %(Eq.~\ref{eq:w1})
and $r_1$ %(Eq.~\ref{eq:r1})
represent kinetics of resource depletion during work and recovery, respectively, and $\delta(t) = 1$ when the user is working on a question and $0$ otherwise.
The precise form of the kinetic functions $w_1$ and $r_1$ were chosen to represent enzyme-catalyzed kinetics with anomalous diffusion (see Methods).

Emerging evidence suggests that glucose may not be the primary energy source for neurons engaged in intensive activity; instead, lactate metabolism may be more important for this function~\cite{Hu:1997vf,Aubert:2005gn,Brown:2007dg,Belanger:2011hy,schurr2011aerobic}.  This motivated us to construct a second model of cognitive depletion. %, specified by Eq.~\ref{eq:2resource}.
This model considers a primary resource $A$, which is responsible for performance but is normally low during rest conditions. Engaging in the task consumes resource $A$, but also causes conversion of a secondary resource $B$ into $A$~\cite{Hu:1997vf,Aubert:2005gn,Cloutier:2009jp,Wyss:2011kt}:
\begin{eqnarray}
\dot{A}(t) &=& -f(A,t) + w_2(A,B,t) \delta(t) \\
\dot{B}(t) &=& -w_2(A,B,t) \delta(t) + r_2(B,t)(1-\delta(t)),
\end{eqnarray}
where $f$ %(Eq.~\ref{eq:f})
is the depletion rate of primary resource $A$,  $w_2$ %(Eq.~\ref{eq:w2})
is the conversion rate of a secondary resource into a primary resource, and $r_2$ %(Eq.~\ref{eq:r2})
is the rate of recovery for secondary resource.

Both models are parameterized by rate constants, which were estimated using model fitting procedure (see Methods) by maximizing the explanatory power of the model, i.e., by maximizing the mutual information between the dynamic values of $A$ or $A,B$ and  the outcome of the corresponding question (correct or incorrect).
Mutual information $MI$ is the reduction in entropy of a random variable achieved by knowing another variable. The advantage of using mutual information, rather than another quantity, in the optimization procedure, is that it does not require knowing the precise way that hypothetical resources translate into performance to find  parameters that maximize the explanatory power of the model.

As mentioned earlier, a significant technical challenge in modeling human data is the large variation among users.   In our sample, users differ substantially in  1) the number of questions they attempt, 2) the length of time they worked without a break and 3) the speed of answering a question.  To handle individual heterogeneity, we characterized each user by two performance-independent parameters.  First, to quantify whether a user was faster or slower than average, we measured the mean time taken by each user to answer a question correctly relative to the population's average time to answer that question correctly. Second, as a proxy for the maximal amount of the available cognitive resources, we measured the longest time each user spent answering a question correctly.  Each user's rates were scaled based on these two numbers, as described in Methods\remove{Sec.~\ref{sec:mm}}.  This user-specific  scaling is based only on the observed time series of user's answers and is not an explicit fitting step nor post-hoc constraint on user performance (beyond knowing they answered at least one question correctly).   This scaling procedure resulted in significant improvement of $MI$ estimates ($>5$ fold improvement) and clearly shows that underlying user heterogeneity must be incorporated in any study of cognitive depletion.
%If experimenters do not seek to control for this variation in user characteristics, they expose themselves to significant sources of both bias and increased variance.

The two-resource model, where the driver of cognitive performance is a primary resource that is converted from a secondary resource, better
links many important performance-related metrics to available resources than the one-resource model.
Table~\ref{tab:mi-results} reports mutual information between actual user performance and resources estimated by the one-resource and two-resource models.
The two-resource model accounts for 12\% of the variance in user's performance on a question,
compared to the one-resource model, which accounts for just 5\% of the variance.
Similarly, the two-resource model explains 16\% of uncertainty in whether or not a user will answer the same question correctly in the future (learning). In contrast, the one-resource model accounts for only 3\% of uncertainty in learning. The two-resource model also explains 16\% of  uncertainty in how long a user will spend between questions and 8\% of how long a user will spend answering a particular question, knowing only the resources available at the beginning of the question, compared to the one-resource model, which explains 4\% and 2\% of the variance respectively.

\begin{figure}[htbp] %  figure placement: here, top, bottom, or page
\centering
\subfloat{\label{fig:speedP}}
\subfloat{\label{fig:speedS}}
\subfloat{\label{fig:learnP}}
\subfloat{\label{fig:learnS}}
\subfloat{\label{fig:perfP}}
\subfloat{\label{fig:perfS}}
\subfloat{\label{fig:speedSec}}
\includegraphics[width=5.5in]{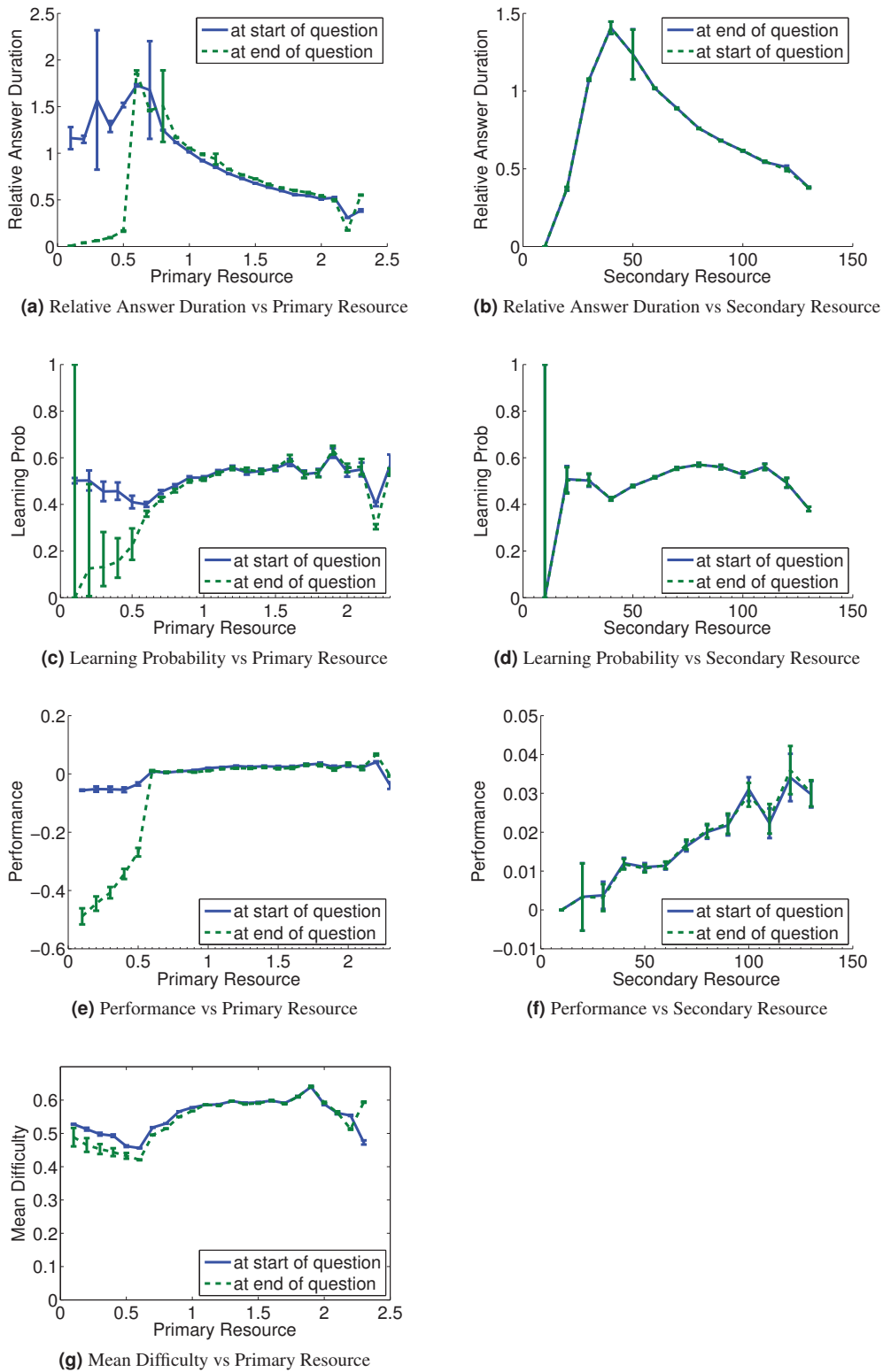}

\caption{\textbf{Key performance measures as a function of estimated cognitive resources.}  Plots show relative answer duration as a function of (a) primary ($P$) and (b) secondary ($S$) resources estimated by the two-resource model.  Similarly, (c), (d) show learning probability and (e), (f) show performance (relative answer accuracy) as a function of estimated primary and secondary resources respectively.  Finally, we also show (g) mean question difficulty vs primary resource.
Error bars show standard errors.
}
    \label{fig:twofactor}
\end{figure}

After optimizing, we used the models to estimate the levels of resources in users at the beginning and end of \added{each question they answered from a time series of their} question-answer attempts.
Figure~\ref{fig:twofactor} shows how various \added{aspects of} performance \deleted{characteristics} depend on the estimated resources of the two-resource model.
\added{The two lines in each subplot correspond to the estimated levels of resources users have when they start working on a question and at the end of the question.}
Answer-speed is determined by both primary and secondary resources (Figs.~\ref{fig:speedP} and~\ref{fig:speedS}).  The more of these users possess, the faster they answer questions.  As resources decline, answer speeds decline, but below a certain threshold users answer questions very rapidly relative to the average  time they spent on the questions. At these levels of the primary resource, relative accuracy also declines rapidly (Figure~\ref{fig:perfP}).
\added{In situations when performance is saturated, differences in resources between the beginning and end are not important; hence, the lines overlap.}
Together, these observations suggest that when \deleted{their primary resources} \added{their primary resources} are too low, \added{they get rapidly depleted over the course of working on questions, leading to decreased accuracy on answers users produce as they}
guess answers. They answer quickly, but at the expense of accuracy. Similarly, when their primarily resources are low at the end of the question (when the correct answer is revealed), users may have more difficulty remembering or learning correct answers to questions, as shown in Figure~\ref{fig:learnP}. \added{This is consistent with primary resource being the driver of performance. }
Note that performance does not suffer as much when primary resource is low at the beginning of the question, which could indicate either depletion or the user starting off in a cold state, i.e., with low initial resources. In the former case, performance will be negatively impacted by depleted resources, while in the latter case, primary resource levels will increase over the course of the question-answer attempt due to conversion of the secondary resource. This will result in improved performance after a brief ``warm-up'' period.

\added{Secondary resources change more slowly, hence, the levels at the beginning (i.e., when users start working on a question) and end of a question are highly correlated and those lines overlap. Low secondary resources constrain primary resources, which tends to lead to lower performance.}

To check whether
%the resource depletion model merely shadows a different, non-resource feature by coincidence,
our findings could be explained by another---non-resource---feature,
we tested alternative hypotheses. A commonly proposed  explanation for performance decrease %assumes
posits that people are sensitive to their near-term success, for example, measured by the fraction of the previous five questions they answered correctly.  When performing successfully, they may be highly motivated and engaged, but when success declines they may become discouraged and inattentive.  Could  the resource model simply  be proxy for the  positive or negative feedback a user receives by answering questions correctly or incorrectly?  To test this hypothesis, we used conditional mutual information, which is defined as $CMI(X,Y|Z) = MI(X,[Y Z]) - MI(X,Z)$.
If variable $Y$ (e.g., current available resources) is merely a proxy for variable $Z$, (e.g., near-term success), then $CMI(X,Y|Z)$ will be smaller than $MI(X,Y)$. In the extreme case where $Y=Z$, $CMI(X,Y|Z)=0$. If, on the other hand, $Y$ and $Z$ together explain $X$ (e.g., performance) better than either alone, than $CMI(X,Y|Z)$ will be larger than $MI(X,Y)$.

We find that the two-resource resource model cannot be explained by near-term success: the $CMI$ between performance and resources conditioned on successfully answering the previous five questions is statistically unchanged from $MI$ between performance and resources, shown in Table~\ref{tab:cmi-alt}.  We also find that the user's parameters used to fit the model do not explain the mutual information between performance and resources,  nor does the difficulty of the question, nor the time spent on the question. We also test if learning can be explained by the time spend on a question or the time until the next question (i.e., potentially forgetting the answer).   We can conclude that the explanatory power of resources is not a simple proxy for any of the alternative explanations. In fact, there is a distinct synergy in knowing both of these features. For example, the intuition that a user may get discouraged by a streak of poor performance, leading them to take a longer break, may \emph{also} be correct, but it does not rule out the resource model.
We also note that results cannot be attributed to random chance (Supplementary Information).

\section{Discussion}
% Results and Discussion can be combined.
Modeling results suggest that performance on practice standardized tests is tied to levels of cognitive resources. As these resources are depleted by sustained mental effort, performance and answer speeds decline, and users have more difficulty learning correct answers.
This is consistent with depletion resulting in mental fatigue, which degrades performance.
In addition, the model suggests that performance may be critically tied to a resource that is normally low when the user isn't on-task, requiring a ``warm-up'' period to raise its levels sufficiently, e.g., solving a simple problem before the start of a test session.
% 2factor vs 1factor
The two-resource model better explains observed performance on test questions than alternative theories. Even in uncontrolled environments of at-home practice tests, it succeeds at explaining over 10\% of the uncertainty in performance. While small in absolute size, this effect is significant to the user, because it changes what is normally a roughly 50:50 odds of answering a question correctly into roughly 70:30 odds. Therefore, it may be possible to improve performance on standardized tests simply by better managing available resources.   Granted, the one-resource model may have a simple parsimonious interpretation: a cognitive resource, specifically glucose~\cite{Gailliot:2007fv,Gailliot:2007dr}, being directly depleted by answering questions. However, it cannot explain as many auxiliary observations nor produce a similar improvement in question answer odds as the two-resource models.
This may explain reported inconsistencies linking ego depletion to levels of glucose in laboratory studies~\cite{Lange:2014iv}. Our findings suggest an alternate mechanism, where the driver for cognitive performance is drawn from a secondary resource. This is consistent with how the lactate shuttle is believed to function in brain metabolism~\cite{Pellerin:2007de}.

Other phenomena not linked to cognitive resources could have similar behavioral outcomes. For example, instead of consuming a resource, cognition could be inhibited by accumulated stress hormones, dopamine, etc.~\cite{Philip:2005jp,Boksem08}. In addition, top-down processes of cost-benefit analysis for continuing work~\cite{Boksem08} could also account for some of our observations.   However, we specifically tested proxies of motivational factors, which did not present significant explanatory power compared to the two-resource model.
Further controlled experiments are required to shed more light on the physiological origin of these phenomena.

Human behavior data are becoming increasingly available, offering new opportunities and new tools for addressing cognitive science questions. When combined with controlled laboratory studies, data analysis promises to accelerate the development and testing of  theories of human behavior and cognitive performance.
The main challenge in making the most of these data is controlling for individual %heterogeneity.
variability. This paper described our solution to this challenge.
By applying these data analysis and modeling techniques to large-scale practice test data, we find that cognitive depletion may explain some of the observed variance in test performance.
By accounting for individual's cognitive resources, we will be better able to predict cognitive performance and devise strategies for improving it.

\section{Methods}\label{sec:mm}

\subsection{Data Availability}

The data were obtained from the Kaggle ``What do you know'' competition ({\tt http://www.kaggle.com/c/WhatDoYouKnow}) provided by grockit.com.  The full data are available at the site, along with the schema.  For the description of the data and how it was used in the competition, see \cite{Grockit1,Grockit2,Grockit3}.

\subsection{Data Processing}
For the purpose of the present work, we utilized the following entries: outcome, user\_id, question\_id, track\_name, round\_started\_at, deactivated\_at.  With this information, we can determine when (round\_started\_at) a user (user\_id) started each question (question\_id) and when they answered it (deactivated\_at) and if they answered correctly, incorrectly, skipped it, or abandoned it (outcome).  For our analysis, we restricted our attention to users who answered at least 15 questions, leaving us with 180 thousand users who answered 6 thousand different questions, for a total of 2.8 million different attempts.  By comparing correct answers with question\_id, we determined that each question\_id has a unique answer.

To determine if a question was a typical math or verbal question, we assigned each track\_name to either math or verbal as follows. Tracks `ACT Math', `ACT Science', `GMAT Quantitative', and `SAT Math' were tagged as math, while the tracks `ACT English', `ACT Reading', `SAT Reading', and `SAT Writing' were tagged as verbal. Unless indicated otherwise, we calculated separate statistics, rates, and outcomes for each user for math and for verbal, based on the obvious observation that people have different competencies.

As is typical of all real-world data, some idiosyncrasies exist in the grockit.com data. For example, sometimes the data indicated that a user answered a question after they started the next question (potentially by navigating back to the previous question).  In this case, we consider the time when the user starts question $i+1$ as the definitive end of their work on question $i$.  When a user abandons a question, e.g. shutting down their computer, stopping work entirely, or timing out on the question, the question is marked in the data as `abandoned.'  We consider the user to be working up to the point at which the data states the user deactivated the question, even if they abandoned the question.  We did not attempt to guess whether the user was actually thinking about the question, nor did we attempt to place any artificial upper bounds on the time a user would spend on a question, to avoid investigator created bias. Given this, we considered the time each user spent on the $i$th question, $T_i$ as
\begin{equation}\label{timeonquestion}
T_i = \min (\text{deactived\_at}_i,\text{round\_started\_at}_{i+1})  - \text{round\_started\_at}_{i}.
\end{equation}

If a question was marked as skipped, this indicates that the user made the affirmative choice to skip a question. Although it is not technically a wrong answer, it is not the correct answer, and it is counted as such unless noted otherwise.

\subsection{Dynamic Models of Resource Depletion}

We explore two dynamic models of cognitive resource depletion. The one-resource model considers a single resource $A$,  that gets depleted while the user is answering test questions and recovers during time periods between question-answer attempts:
\begin{equation}\label{eq:1resource}
\dot{A}(t) = -w_1(A,t)\delta(t) + r_1(A,t) (1-\delta(t)),
\end{equation}
where $w_1$ is a function representing kinetics of depletion during work, $r_1$ is the kinetics of recovery, and $\delta(t) = 1$ when the user is recorded as working on a question and $0$ otherwise.
The precise form of the kinetic functions were chosen to represent enzyme-catalyzed (Michaelis-Menten) reactions with anomalous diffusion~\cite{Xu:2007ca}. % under heterogeneous conditions~\cite{Xu:2007ca}.
Specifically, we use the following expressions for rates:
\begin{eqnarray}
\label{eq:w1}
w_1(A,t)&=&\frac{k}{t^\rho} \frac{A(t)}{(K_m + A(t))} \\
\label{eq:r1}
r_1(A,t)&=& \frac{k_r}{t^\rho} (A_{max} - A(t)),
\end{eqnarray}
where $K_m$ is the Michaelis constant, %which is the substrate concentration at which the reaction rate is half of {\displaystyle V_{\max }} V_{\max }.[1]
$k_w$ and $k_r$ represent the forward and reverse chemical reaction rates,
and $\rho$ is the exponent characterizing anomalous diffusion. In the equation above, $A_{max}$ represents the maximum amount of resource $A$.

Emerging evidence suggests that glucose may not be the primary energy source for neurons engaged in intensive activity; instead, lactate metabolism may be more important for this function~\cite{Hu:1997vf,Aubert:2005gn,Brown:2007dg,Belanger:2011hy,schurr2011aerobic}.  This motivated us to construct a second model of cognitive depletion.
We compare the model above to a second model of cognitive depletion.
This two-resource model considers a primary resource $A$ and a secondary resource $B$. Engaging in the task consumes resource $A$, but also causes conversion of a secondary resource $B$ into $A$:%~\cite{Hu:1997vf,Aubert:2005gn,Cloutier:2009jp,Wyss:2011kt}:
\begin{eqnarray}
\label{eq:2resource}
\dot{A}(t) &=& -f(A,t) + w_2(A,B,t) \delta(t) \\
\dot{B}(t) &=& -w_2(A,B,t) \delta(t) + r_2(B,t)(1-\delta(t)),
\end{eqnarray}
where $f$ is the rate of consumption of primary resource $A$,  $w_2$ is the conversion rate of a secondary resource into a primary resource, and $r_2$ is the rate of recovery for secondary resource.
The functions in the two-resource model are:
\begin{eqnarray}
\label{eq:f}
 f(A,t) &=& \frac{k_w}{t^\rho} \frac{A(t)}{K_A + A(t)}  \\
 \label{eq:w2}
 w_2(A,B,t) & \equiv &\frac{k_b}{t^\rho} \frac{(1 - A(t))B(t)}{K_B + B(t)}\\
  \label{eq:r2}
 r_2(B,t) &=&  \frac{k_r}{t^\rho} (B_{max} - B(t)) (1-\delta(t)).
\end{eqnarray}
The functional form of the parameters, although nontrivial, is a natural extension of the one-resource model, allowing for complex and anomalous enzyme kinetics.

\subsection{User Characterization}

To account for user heterogeneity, we decided to fit one set of kinetic parameters, but each parameter was scaled for each user based on performance-independent observable user behavior. First, we measured the longest time it took each user to answer a question correctly,
\begin{equation}
T_L = \max_{i \in correct} T_i,
\end{equation}
where $T_i$ is the time the user spends answering question $i$, determined as in Eq~\ref{timeonquestion}.  This restricts our investigation to  users who answered at least one question correctly.
%Although the user may have answered the question correctly purely by chance,
Although the user may sometimes get the right answer purely by chance,
we hypothesize that the user will only be able to answer questions correctly when they have sufficient cognitive resources and that the longest time they spend answering a question correctly will scale with the total depth of their cognitive resources.  Second, we measured for each user the average time it took them to answer a question correctly, relative to the average time it took all users to answer that question correctly.  More specifically,
\begin{equation}
T_r = \frac{1}{N_c} \sum_{i \in correct} T_i/\langle T_i \rangle,
\end{equation}
where $N_c$ is the number of questions the user answered correctly, $T_i$ is the time the user spent on the $i$th question, and $\langle T_i \rangle$ is the mean time \emph{all} users took to answer that same question correctly. Thus, $T_r$ reflects if a user is faster than average or slower than average when attempting to answer a question.  We only consider correctly answered questions to remove the effects of skipping, guessing (most guesses will be incorrect), and abandoning.

User parameters were constrained to fall between the 5th and 95th percentiles, which were 33s and 200s respectively for math $T_L$, 29s and 240s for verbal $T_L$, 0.46 and 1.6 for math $T_r$, and 0.45 and 1.7 for verbal $T_r$.  Parameters falling below this range were automatically set to the 5th percentile for that value, while those falling above the range were set to the 95th percentile. This procedure removes pathological effects due to users who only guess, users who answer questions abnormally slow, etc.

\begin{figure}
\centering
\includegraphics[width=\figwidthW]{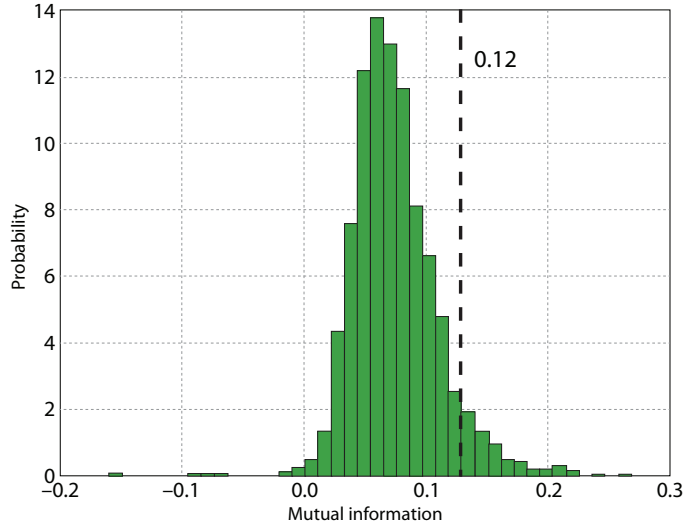}
\caption{Fitting separate rates for each user produces a lower mean mutual information than the mean mutual information produced by having a single set of free parameters combined with user-specific scaling, as described in the text.  The dashed line shows the mean mutual information for the user-specific scaling results, 0.12. \label{fig:comparison} Having 5 parameters per user results in significant overfitting. As described in the text, user-specific scaling is not a fitted free parameter.}
\end{figure}

For the fitting procedure (detailed below), we scaled all of the users rates as $k\rightarrow k/T_r$, where $T_r$ is specific to each user.  In addition, we scaled $B_{max} \rightarrow B_{max}f_0/T_r,$ where %$f_0 =  \{ \log(T_L + 1) \text{ if } \rho = 1, \text{ or } (T_L+1)^{-\rho} ((T_L+1)^{\rho} - T_L - 1)/(\rho -1)\}.$
\begin{equation}
%\[
    f_0=
\begin{cases}
   \log(T_L + 1),& \text{if } \rho = 1\\
    \frac{((T_L+1)^{\rho} - (T_L + 1))}{(T_L+1)^{\rho}(\rho -1)},              & \text{otherwise}
\end{cases}
%\]
\end{equation}
In this way, rates for faster users who were able to work longer successfully were different from slow users or users who could not maintain high performance levels over extended periods. We also tried fitting each user separately, where each user's parameters was determined only from that user's data, but we found this to be inferior to the results from user-specific scaling Table~\ref{tab:mi-results}. That is, we find having 5 free-parameters and user-specific scaling produced superior results than fitting 5 free-parameters \emph{per user}, shown in Fig.~\ref{fig:comparison}. This is most likely due to increased overfitting. To carry-out a test train split for user-specific rate, we train on the first half of each user's time series and test on the second half. This is less robust than fitting on one set of users and testing on another set of users, and more prone to over-fitting.  However, there is no alternative if you want to do user-specific rates.

% show time series of work/rest for a user?
\subsection{Mutual Information}
Mutual information ($MI$) is the reduction in entropy (uncertainty) of random variable $X$ achieved by knowing another variable $Y$. For example, how much information about an individual's performance ($X$) do we obtain by knowing the amount of cognitive resources available to him or her ($Y$)? $MI$ is defined as $MI(X,Y) = H(X) - H(X|Y)$, where $H(X)$ is the entropy of the random variable $X$, and $H(X|Y)$ is the entropy of $X$ given $Y$, and it is measured in bits.  Mutual information has the property that $MI(X,Y) = MI(f(X),g(Y))$ for invertible functions $f$ and $g$~\cite{Nair:2006ws}, so it is not necessary to know the precise way that %hypothetical
hypothesized resources translate into performance to find  parameters that maximize the explanatory power of the proposed models. In contrast, optimizing a regression model or Pearson correlation for  parameter estimation requires not only correctly modeling resources but also knowing how those resources quantitatively translate into predicted performance, because $R^2$ is only maximized when predicted and observed results have an affine relation.
Contrast this to $MI$, where a large $MI$ implies large explanatory power, even if we do not know the precise mapping from $X$ to $Y$.

To calculate mutual information, we utilized a variation of the method described in~\cite{pal2010estimation}.  That is, to calculate $MI(X,Y)$, we first performed a copula transform on each dimension of $X$ and $Y$.  We then calculated the quantity,
\begin{equation}
MI(X,Y) =  - H([X Y])  + \frac{1}{10} \sum_{i=1}^{10} H([X \tilde{Y}_i]) + H([\tilde{X}_i Y]) - H([\tilde{X}_i \tilde{Y}_i]),
\end{equation}
where $H([X Y])$ is the Renyi entropy with Renyi-exponent taken as 0.99999 (that is, a very close approximation to Shannon entropy), calculated according to Pal et al.~\cite{pal2010estimation}. $\tilde{X}_i$ is a shuffled version of $X$, such that any correlation between dimensions of $X$ or $Y$ are destroyed, and $i$ denotes the $i$th shuffle.  This version of mutual information was chosen because it demonstrated the most robustness.  Normal calculations of mutual information, calculated directly as $H(X) + H(Y) - H([X Y])$ subtract entropy calculated from  low-dimensionality spaces from entropy calculated from high-dimensional spaces.  Because bias in entropy calculations vary with dimension, the shuffled version helps to cancel out systematic dimensional biases.

\subsection{Model Fitting}

To determine the parameters to use for estimating resource levels, we first divided the data into a training and test sets. The training set comprised of 250 users, each making at least 500 attempts at one or more question.  Out of the training set, we only measured the performance between the 2nd attempt and the 5000th attempt (should the user make more than 5000 attempts), to avoid overweighting the statistics with attempts from a small handful of users.  The test set comprised  users who made at least 25 attempts, answered at least 20\% of the question correctly, \emph{and} were not included in the training set.  The training set was used to find parameters, and the results in the tables, figures and tables were all produced from the test set. As stated in the section on user characterization, the global rates were scaled for each user according to performance-independent observations.

To find parameters, for each user in the training set, we evaluated their resources at the beginning and end of each attempt
using Eqs.~\ref{eq:1resource} or \ref{eq:2resource}.
As mentioned above, we calculated resources for math and verbal questions separately, so we really calculate two different independent sets of resources for each user.  When a user is working on a math question, we consider that to be recovering for the verbal resources, and vice versa.  For the purpose of comparing resources to performance, we only compare the resource that matches the type of question.
So, if the user is attempting to answer a math question, we compare the outcome with the resources connected to math.
%So, if the current question is a math question, we compare the outcome of that question with the math resource.

We then optimized the parameters using the NLopt optimization library~\cite{NLopt} with COBYLA algorithm~\cite{COBYLA}.  The resulting parameters were all constrained to be between 0.0001 and 2.0.  The results of the fitting were as follows:
for the one factor model, $k=0.078,k_r=0.21,\rho=1.0,K_m =0.44867$, and for the two factor model, $k_w = 0.003,k_b = 0.118,k_r = 0.00125,B_{max}=0.27,\rho = 0.03$.  In addition, we fixed $K_A=0.858$ and $K_B=0.1$, which were values taken from~\cite{Cloutier:2009jp}.

\subsection{Odds Adjustment}\label{sec:odds}
Entropy for a binary outcome (correct or incorrect) is defined as $S = -p\log_2(p) - (1-p)\log_2(1-p),$ where $p$ is the probability of answering the question correctly.  For all users, averaging over all questions, the total entropy for getting the question correct or not is roughly 1 bit, meaning without knowing anything about the user or the question, the user has a roughly 50:50 chance of getting the question correct.  As we report in the main text, the two resource model accounts for roughly 12\% of performance variance, so once you know the resources at the beginning of the question, the remaining entropy is 0.88 bits. Thus, we have $0.88 = -p^*\log_2(p^*) - (1-p^*)\log_2(1-p^*),$ where $p^*$ is the probability of answering the question correctly (or incorrectly).  Thus, $p*$ is  0.7 or 0.3.
Therefore, knowing the resources, or estimating them using the model,  changes the odds the user answers that question correctly 50:50 to 70:30 (high resources) or 30:70 (low resources), even without incorporating anything else about the question or user.  The user may then  make adjustments in their test-taking strategy to improve the overall score.

\section*{Acknowledgements}
This work was supported, in part, by AFOSR (contract FA9550-10-1-0569), by DARPA (contract W911NF-12-1-0034), by ARO (contract W911NF-15-1-0142) and IARPA (contract 2017-17042800005). The research described in this paper is also part of the Analysis In Motion Initiative at Pacific Northwest National Laboratory. It was conducted under the Laboratory Directed Research and Development Program at PNNL, a multiprogram national laboratory operated by Battelle for the U.S. Department of Energy.

% Either type in your references using
% \begin{thebibliography}{}
% Text
% \end{thebibliography}
%
% or
%
% Compile your BiBTeX database using our plos2015.bst
% style file and paste the contents of your .bbl file
% here.
%
%\bibliographystyle{spmpsci}
%\bibliography{references,references2}

%\section{Author contributions statement}
%NH and KL designed the study and NH, JH, and SY analyzed data. NH, JH, SY and KL contributed to the manuscript.
%
%\section{Additional information}
%
%\noindent \textbf{Competing financial interests:} Authors declare no competing financial interests.
%
%\noindent \textbf{Ethics:} University of Southern California's and PNNL's Institutional Review Boards (IRB) classified research as ``non-human subjects research.''
%

\section*{Figures}

\section*{Tables}

\begin{table}[htbp]
   \centering
  {} % requires the topcapt package
   \begin{tabular}{@{}|l|ccc|ccc| @{}}
      \hline
        &  \multicolumn{3}{|c|}{Two-Resource Model} &\multicolumn{3}{c|}{One-Resource Model}  \\
      \cmidrule(lr){2-4} \cmidrule(lr){5-7} % Partial rule. (r) trims the line a little bit on the right; (l) & (lr) also possible
      \small{Quantity}    & \small{Bits} & \small{Std. Dev.} & \small{\% of Entropy} & \small{Bits} & \small{Std. Dev.} & \small{\% of Entropy}\\
      \hline %\midrule
      MI(A;R)      & 0.12 & 0.015 &  12\%  & 0.04 & 0.006 & 4\%\\
      MI(L;R)         & 0.10     &  0.02  & 16\% & 0.03 &  0.014 & 3\%\\
      MI(T;R$_b$)   & 0.51  & 0.05 &  8\%  & 0.15 & 0.01 & 2\% \\
      MI($\Delta$T;R) & 1.18   & 0.06 & 16\% & 0.27 & 0.016 & 4\% \\
      \hline %\bottomrule
   \end{tabular}
   \caption{\textbf{Model comparison using Mutual Information.}
   %Mutual Information versus estimated cognitive resources provides a measure for explanatory power of cognitive depletion model.}
   Mutual Information ($MI$) between observed performance and cognitive resources $R$, estimated by a model, provides a measure of the explanatory power of the model. Performance variables are $A$ (questions answered correctly), $L$ (learned correctly), $\Delta T$ (time until next question), $T$ (time spent on question).  The variable $R$ represents resources at the beginning and end of question,  while $R_b$  specifically represents resources at beginning of question.
   The two-resource model better explains observations, capturing more information about them (in bits) and explaining more of the entropy.
   }
   \label{tab:mi-results}
\end{table}

\begin{table}[htbp]
   \centering
   \begin{tabular}{@{}|l|cc| @{}}
      \hline %\toprule
      \textbf{Alternative Hypothesis}    & \textbf{Bits} & \textbf{Std. Dev.}\\
      \hline %\midrule
      CMI(L;R$|\Delta$T)    & 0.14 & 0.03\\
      CMI(L;R$|$T)    & 0.10 & 0.03\\
      CMI(A;R$|$U$_5$)       & 0.11     &  0.02  \\
      CMI(A;R$|$P)   & 0.18 & 0.02 \\
      CMI(A;R$|$T) & 0.14 & 0.02\\
      CMI(A;R$|$D) & 0.14 & 0.02\\
      CMI($\Delta$T;R$|$U$_5$) & 1.36 & 0.06\\
      \bottomrule
   \end{tabular}
   \caption{\textbf{Comparison of the explanatory power of alternative hypotheses.} Conditional Mutual Information ($CMI$) gives mutual information between observed performance and cognitive resources $R$, conditioned on the operative variable of the alternative hypothesis.  Performance variables are $A$ (questions answered correctly), $L$ (learned correctly), and time  $\Delta T$ until next question,. Alternative hypothesis variables are time spent on the question ($T$); average performance on previous five questions ($U_5$), parameters defining user ($P$), and difficulty of the question ($D$).
      Alternative hypotheses do not rule out cognitive depletion,
   because $CMI$ between performance and resources conditioned on alternative hypothesis is statistically unchanged. For example, conditioning on successfully answering the previous five questions ($U_5$) does not change mutual information between resources and performance.
   }
   \label{tab:cmi-alt}
\end{table}

\section*{Additional Files}
  \subsection*{Additional file 1 --- Supplementary Material}
  Additional regression fits and results of diagnostic tests.

\end{document}